\documentstyle[epsf]{mn}

\begin{document}

\title[EUV and X-ray variability of NGC~4051]
{Simultaneous EUV and X-ray variability of NGC~4051}
\author[P. Uttley et al.]
{P. Uttley$^{1}$\thanks{e-mail: pu@astro.soton.ac.uk}, 
I. M. M$^{\rm c}$Hardy$^{1}$, I. E. Papadakis$^{2}$,
I. Cagnoni$^{3}$, A. Fruscione$^{4}$  \\
$^{1}$Department of Physics and Astronomy, University of Southampton, 
Southampton SO17 1BJ \\
$^{2}$Physics Department, University of Crete, PO Box 2208, 710 03
Heraklion, Crete, Greece \\
$^{3}$SISSA-ISAS, Via Beirut 4 - 34014, Trieste, Italy \\
$^{4}$Harvard-Smithsonian Center for Astrophysics, 60 Garden Street,
Cambridge, MA 02138, USA 
}

\date{Accepted....  Received 1998 November 5; in original form
1998 November 5}

\maketitle
\parindent 18pt
\begin{abstract} 
We present a flux variability study of simultaneous {\it RXTE} and {\it
EUVE} observations of the highly variable Seyfert galaxy NGC~4051. 
We find a strong correlation between variability in the EUV
and medium energy X-ray bands,
indicating that both are sampling the same power-law continuum.  The lag
between the two bands is less than 20~ks and, depending on model
assumptions, may be $<1$~ks. 
We examine the consequences of such a small lag in the
context of simple Comptonisation models for the production of the power-law
continuum.  A lag of $<1$~ks implies that
the size of the Comptonising
region is less than 20 Schwarzschild radii for a black hole of mass 
$>10^6$~M$_\odot$.      
\end{abstract}


\section{Introduction}
Despite two decades of observations, the mechanism for the production of
the X-ray continuum of radio-quiet AGN remains poorly
understood.  The power-law nature of the continuum implies a
Comptonisation origin, possibly upscattering of low energy (optical, UV)
photons by a thermal distribution of energetic electrons (as indicated
by the OSSE detection of high energy cutoffs in Seyfert spectra e.g.,
Zdziarski et al., 1997).  Rapid variability of the continuum
implies that the X-rays are produced in a small region, close to the
putative central black hole.  These conclusions are still vague;
further understanding of the innermost regions of AGN requires more
detailed knowledge of the size and geometry of the X-ray producing
region, the source of the low energy `seed' photons and the physical
characteristics of the scattering particles.  \\
We can begin to address these points by answering two questions: 

1.  How low (in energy) does the power-law continuum extend? 

2.  Are there any lags between the variations in
the X-ray band and lower energy bands?

Answering the first question will allow us to place an upper limit on
the energy of
the seed photons (and hence allow us to constrain the temperature of
the source of the seed photons).  Answering the second question will
allow us to determine the size and possibly the geometry of the emitting
region.  In upscattering of low energy photons by a
cloud of Comptonising
particles, more scatterings are required to produce high energy photons
than lower energy photons,
hence there will be an intrinsic lag between high and low energy bands. 
The size of this lag is determined by the optical depth and size of the
Comptonising region. 

Searches for lags between the X-ray band and lower-energy contnuum bands
in AGN have proved difficult, due to the difficulty of scheduling simultaneous
observations between space-based and ground-based instruments for the
long durations required to search for correlated variability.  In this
respect, the best studied radio-quiet AGN to date are the Seyfert
galaxies NGC~4151, NGC~5548 and NGC~7469.  The X-ray/UV monitoring
of NGC~4151 and NGC~5548 constrained any lags between the UV
and X-ray continua to be less than 0.15~d (Edelson et al. 1996) and less
than 6~d
(Clavel et al. 1992) respectively.  The intense, month-long monitoring
of NGC~7469 with the Rossi X-ray Timing Explorer ({\it RXTE}) and {\it
IUE} (Nandra et al. 1998) revealed a complex correlated behaviour.  In
this case, the UV continuum led the X-rays by $\sim4$~d at the peaks in the 
lightcurves,
while there was no apparent lag between both bands during the lightcurve
minima, suggesting that complex, multi-zone Comptonisation models may be
required to explain the relationship between the X-ray and UV continua
in NGC~7469.  The similarity between X-ray spectral shape and
variability properties
(e.g. the power spectrum, M$^{\rm c}$Hardy 1988, Edelson \& Nandra 1998)
of Galactic
Black Hole Candidates (GBHCs) and AGN suggests that GBHCs may be
scaled-down analogs of AGN.  A lag of $\sim6$~ms has been measured
between medium-energy (5--14~keV) and low-energy (2--5~keV) X-ray bands
in the GBHC Cygnus X-1 (Page 1985), which scales to a lag of order
$100$~s--10~ks in AGN, if we naively scale by the black hole mass
(for a range of black hole masses, $10^{6}--10^{8}$~M$_{\odot}$, as
measured by Wandel Peterson \& Malkan 1999).  In
simple Comptonisation models, we expect lags between bands to be energy
dependent, with greater lags being seen between bands which are more
separated in energy.  This expected dependence has been seen in GBHCs
(e.g. Cygnus X-1, Nowak et al. 1999), so that if the same applies to
AGN, we might expect lags between the X-ray and EUV bands
to be less than those between the X-ray and UV bands (i.e. less than 1
day) and more easily measurable by a relatively short monitoring
campaign.

NGC~4051 is a nearby ($z=0.0023$), low luminosity
($L_{\rm X-ray}<10^{42}$~ergs~s$^{-1}$) Narrow Line Seyfert~1 galaxy, which
displays strong X-ray variability in both medium energy X-rays \cite
{pl} and low energy X-rays (Lawrence et al. 1987, M$^{\rm c}$Hardy et
al. 1995) with a doubling timescale of
less than 1~ks.  This strong variability makes
NGC~4051 an ideal candidate for searching for lags between the X-ray
band and lower-energy bands.  We therefore obtained
simultaneous observations of NGC~4051 with {\it RXTE} and the
Extreme Ultraviolet Explorer ({\it EUVE}) satellite in May and December
1996.  In this paper we present the
results of these observations as they relate to flux variability. 
Detailed spectral analysis of the {\it RXTE} observations is left to a
later work.

In the next section we discuss the observations and data reduction method,
while in section 3 we present the lightcurves and their cross-correlation
function.  In section 4 we further constrain any lag by
assuming a simple spectral model.  Finally, we discuss the implications of
our results for the size of the scattering region in NGC~4051 and for
Comptonisation models of the X-ray continuum. 

\section{Observations and Data Reduction}
\begin{figure*}
\begin{center}
{\epsfxsize 0.7\hsize
 \leavevmode
 \epsffile{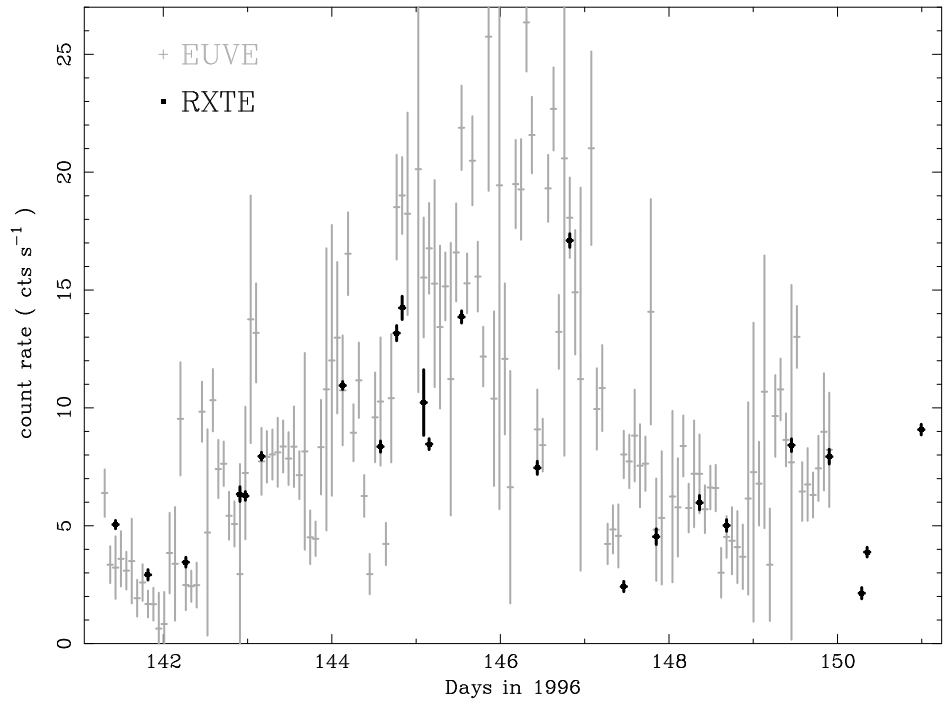}   
}\caption{May 1996: 5544~s binned {\it EUVE} (scaled by factor
100) and {\it RXTE} (3~PCUs, 2--10~keV) lightcurves of NGC~4051 (errors
are $1\sigma$).}
\end{center}
\label{fig:maylc}
\end{figure*}
From May 20 to May 28 1996 {\it RXTE}
observations of typical duration $\sim1$~ks were carried out approximately
every half-day and continuous (subject to source visibility) {\it EUVE}
observations were also carried out at this time.  Additional
simultaneous observations occurred on May 6 1996.  From December 13 to
December 16, a long
{\it RXTE} observation was carried out, which
overlapped with a continuous three-day {\it EUVE} observation for
$\sim120$~ks. 

The {\it RXTE} satellite observed NGC~4051
with the Proportional Counter Array (PCA) and the High Energy X-ray
Timing Experiment (HEXTE) instruments.  The PCA consists of 5 Xenon
Proportional Counter Units (PCUs), sensitive to X-ray energies from
2--60~keV.  The HEXTE covers a range of between 20--200~keV, but due to the
faint nature of the source we only consider the PCA data in this
work. 

Discharge problems mean that of the 5 PCUs
in the PCA, PCUs 3 and 4 are often switched off, so we include data from
PCUs 0, 1 and 2 only.  We extract data from the top layer of
the PCA using the standard {\sc
ftools 4.1} package, excluding data obtained within and up to 20 minutes
after
SAA maximum and data obtained with earth elevation $<10^{\circ}$.  We
generate background data for the PCA with {\sc pcabackest v2.0c} using
the new L7 model for faint sources.   

NGC~4051 was observed with the Deep Survey Spectrometer (DS/S) on
board the {\it EUVE} satellite.  The DS/S (e.g. Welsh et al. 1990) is
equipped with a broad band imaging detector (covering the
66--178~\AA~band in the Lexan/B filter) and three spectrometers
covering the `short' (SW:~70--190~\AA), `medium' (MW:~140--380~\AA)
and `long' (LW:~280--760~\AA) EUV wavelengths (Abbott et al. 1997).
This configuration allows simultaneous imaging and spectroscopy with a
spatial resolution of $\sim 1$ arcmin and a spectral resolution of
$\lambda/\Delta\lambda \sim 200$ at the short wavelengths.

NGC~4051 was detected only in the SW spectrometer because the
interstellar medium ($N_{\rm H}=1.31\times 10^{20}$~cm$^{-2}$ Elvis, Lockman
\& Wilkes 1989) severely attenuates the EUV spectrum for wavelengths
longer than $\sim 100$~\AA. But despite the long ($\sim 203$~ks)
total effective exposure, the average signal-to-noise ratio (SNR)
achieved in the spectrum -- per 1~\AA~bin -- is very low ($\sim1.5\sigma$ in
the range 80-90 \AA) and spectral data will not be discussed further here. 
For the purposes of this work, we only use the lightcurves produced by
the DS imaging detector in the energy range 124--188~eV (with the
Lexan/B filter). 

We extracted the light curves from the DS time-ordered event list
using the {\it EUVE} Guest Observer Center software (IRAF/EUV package) and
other IRAF timing tasks adapted for {\it EUVE} data.  After correcting the
data for instrumental deadtime, telemetry saturation, vignetting and
eliminating intervals of high particle background, the total effective
exposure is 178738~s.  We counted the source photons in a circle of
$2^\prime$ radius, and we estimated the background in a concentric
annulus with inner and outer radii of $2.7^\prime $ and $8^\prime$
respectively. The extraction region includes more than 98\% of the DS
point spread function (Sirk et al. 1997) and the background is
generally very uniform in the chosen area.  The effective area of the
DS instrument (25~cm$^2$ at $\lambda=85$~\AA, Sirk et al. 1997), is
more than 10 times larger than the spectrometer effective area at this
wavelength and a good detection of the source was obtained during 
each 5544~s {\it EUVE} orbit (see figures 1 and 2).
A complete description of the {\it EUVE}
observation and data analysis can be found in Fruscione et al. (1999).
Similar lightcurve extractions have been performed on other AGNs
detected by {\it EUVE} (e.g. 3C273, Ramos et al. 1997, NGC~5548, Marshall et
al. 1997).

\section{timing analysis}      
\begin{figure*}
\begin{center}
{\epsfxsize 0.7\hsize
 \leavevmode
 \epsffile{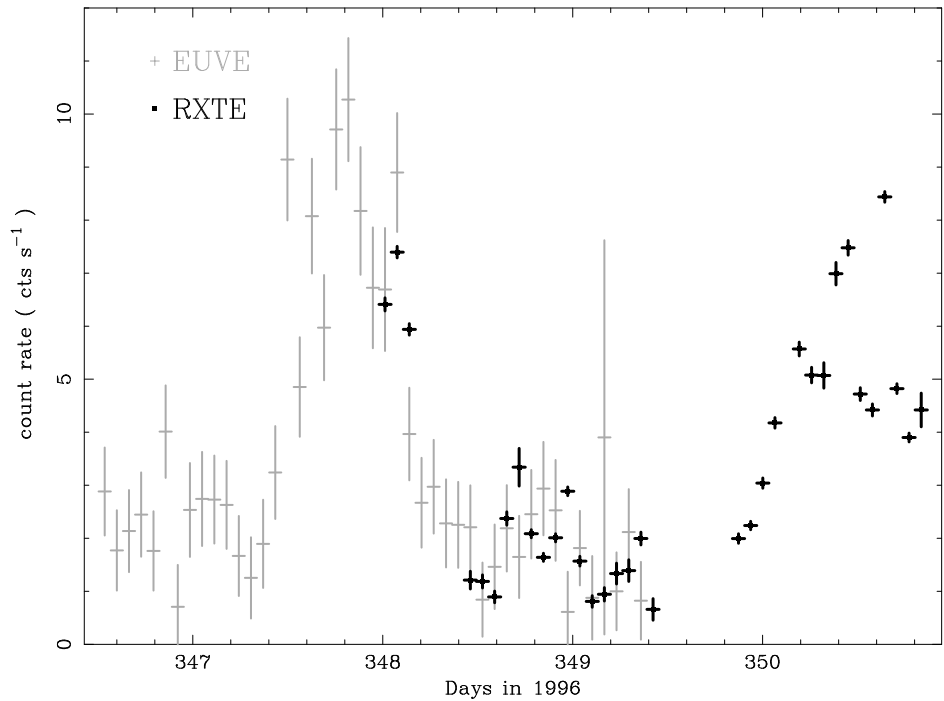}  
}\caption{December 1996: 5544~s binned {\it EUVE} (scaled by
factor 100) and {\it RXTE} (3~PCUs, 2--10~keV) lightcurves of NGC~4051
(errors are $1\sigma$).}
\end{center}
\label{fig:declc}
\end{figure*}
We show the background-subtracted {\it RXTE} (2--10~keV) and
{\it EUVE} lightcurves for the May observations in
Fig. 1.  For presentation purposes, we have scaled up the {\it EUVE}
lightcurve by a factor of 100.  The bin width is 5544~s, which is the
orbital period of {\it EUVE}.  For clarity we do not show the May
6th observations (but we note that the 2--10~keV and {\it EUVE} count
rates for this time were $\sim8.0$~cts~s$^{-1}$ and 0.12~cts~s$^{-1}$
respectively).  As can be seen from the figure, the {\it EUVE} count
rate varies strongly between less than 0.05~cts~s$^{-1}$ and
$\sim0.25$~cts~s$^{-1}$.  The 2--10~keV count rate also varies greatly
between {\it RXTE} observations, with a range of between 3~cts~s$^{-1}$ and
17.4~cts~s$^{-1}$.  The lightcurves for the December observations are
shown in Fig. 2, using the same bin width and the same scaling factor
for the {\it EUVE} lighcurve including, for completeness, the data
obtained by both instruments outside the times of overlap.  Strong
variability can be seen, with NGC~4051 reaching a particularly low
flux state during the period of overlap. 
\begin{figure*}
\begin{center}
{\epsfxsize 0.7\hsize
 \leavevmode
 \epsffile{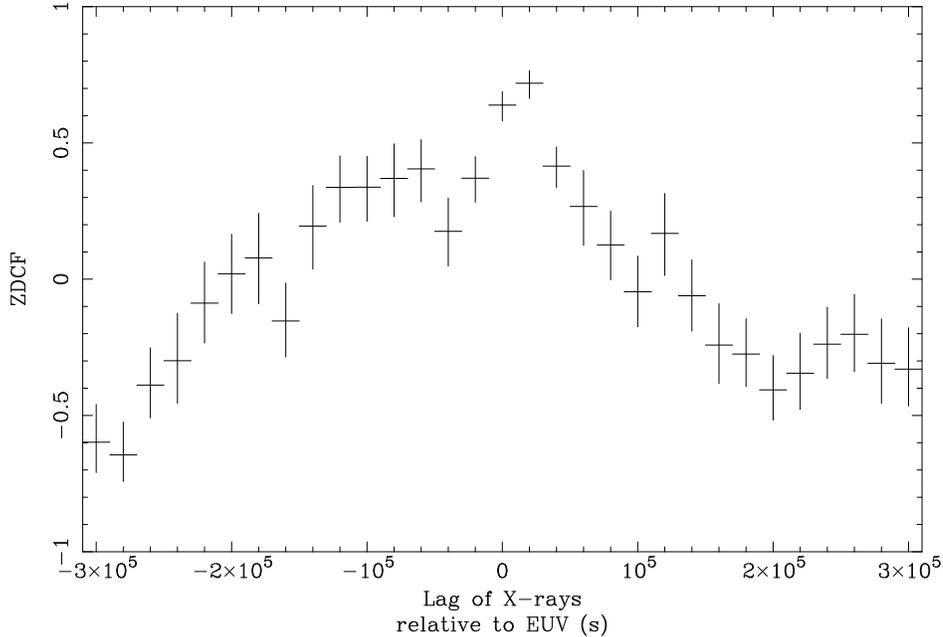}
}\caption{Z-transformed discrete cross-correlation function of {\it EUVE}
and {\it RXTE} (4--10~keV) lightcurves, combined from separate May and December
ZDCFs.  The ZDCF is binned to 20~ks and errors are $1\sigma$.}
\end{center}
\label{fig:combzdcf}
\end{figure*}
Visual inspection of the lightcurves shows a striking
correlation between the two bands.  We can quantify this correlation and
search for lags between the two bands by carrying out a
cross-correlation analysis of the lightcurves.  In the context of this
paper we are particularly interested in the relation of the
power-law component of the X-ray spectrum to the EUV emission.  Since
NGC~4051 is known to display both variable low energy X-ray absorption
and an iron fluorescent emission line at $\sim6$~keV \cite{gu}, we will
exclude these components and obtain the `pure' power-law contribution to
the emission by extracting a lightcurve in the 4--10~keV band, excluding
5--7~keV (corresponding to PCA channels 12--14 and 21--28 in the current
gain epoch 3). 

We compute the cross-correlation function (CCF) of the time series
using the Z-transformed Discrete Correlation Function (ZDCF) method of
Alexander (1997), which is based on the DCF method of Edelson \& Krolik
(1988) but estimates errors more reliably. 
We calculate seperate ZDCFs for May and December 1996 and
bin up the resulting noisy ZDCFs (which have different binning) into
identical 20-ks-wide bins, before
combining them by adding together values corresponding to the same lag
(weighting according to errors).  We show the resulting combined ZDCF in
Fig. 3.

A simple visual inspection of the ZDCF indicates that the EUV band leads
the X-rays by between 0 and 20~ks.  Simulations of perfectly correlated
lightcurves (with zero lag)
which use the same sampling pattern as our data show that 
the peak at
20~ks is probably artificial, due to an excess of data pairs sampled at
that particular lag.  Therefore, we believe that the most likely lag is
within the 0~ks bin.  We now turn to an alternative technique for
constraining the lag, which is not affected by the sampling pattern.

\section{a simple spectral model}
The simplest explanation for the strong correlation between the {\it
EUVE} and {\it RXTE} lightcurves is that both instruments are sampling
the same continuum, i.e {\it the X-ray power-law extends to the EUV
band}.  A detailed spectral analysis of the {\it RXTE} data will be
described in a later work, but here we can test
the hypothesis that the EUV continuum is an extension of the X-ray
power-law continuum by fitting a simple power-law to the 4--10~keV
(excluding 5--7~keV) region of the spectrum for the brightest
observation in May.  We find a power law slope (photon index, $\Gamma$)
of $2.3\pm0.1$, corresponding
to a 2--10~keV flux of $6.5\times10^{-11}$~ergs~cm$^{-2}$~s$^{-1}$.  If
we use these parameters in the {\sc pimms v2.3} count rate calculator,
assuming a galactic absorption of $1.31\times10^{20}$~cm$^{-2}$, we
obtain a predicted {\it EUVE} DS/S (Lexan/B) count rate of
$0.27\pm0.1$~cts~s$^{-1}$.  This agrees well with the actual
{\it EUVE} count rate (simultaneous with the 500~s long {\it RXTE}
observation) of $0.24\pm0.05$~cts~s$^{-1}$.  Note that because the {\it
EUVE} count rate can be adequately explained as the extrapolation of the
X-ray power law modified by Galactic absorption, there is no
requirement for a significant neutral column in the AGN host galaxy, a
result consistent with previous {\it ROSAT} and {\it ASCA} observations.
 The lack of any significant absorbing column (in addition to Galactic)
implies that the significant low energy X-ray absorption seen in
previous observations of
NGC~4051 (M$^{\rm c}$Hardy et al., 1995; Guainazzi et al., 1996)
must be due to ionised gas (so that hydrogen and helium are
virtually completely ionised).  This is again consistent with previous
observations, which indicate that the X-ray absorption is due to an
ionised `warm absorber'. 
\begin{figure*}
\begin{center}
{\epsfxsize 0.7\hsize
 \leavevmode
 \epsffile{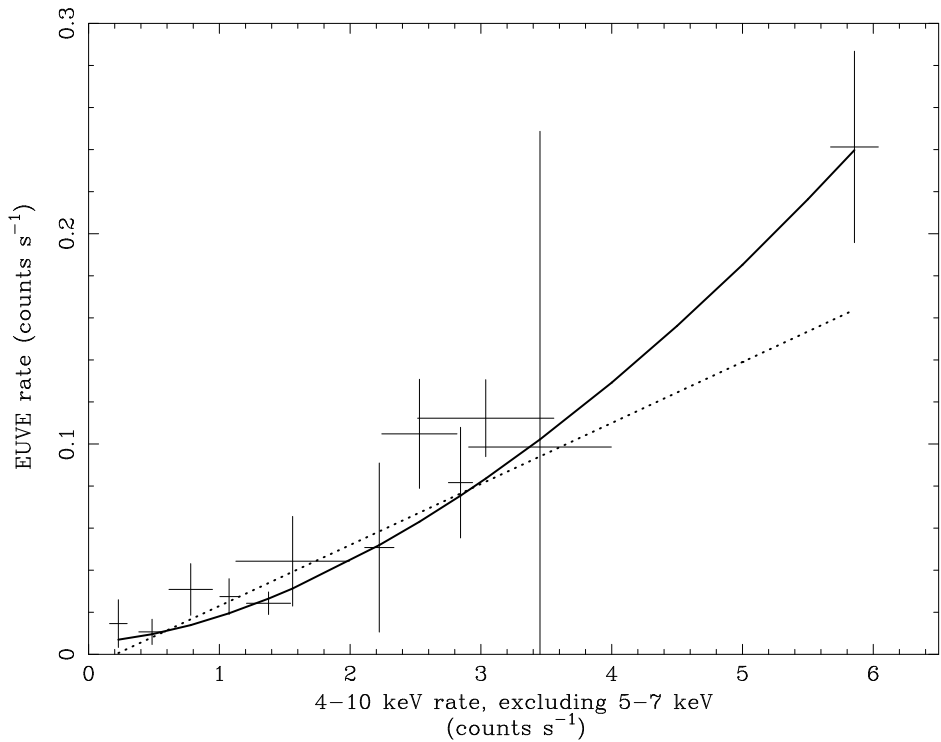}
}\caption{Comparison of linear and non-linear models
(dotted line and solid line respectively).  For clarity the data have  
been averaged (and $1\sigma$ errors calculated accordingly)
into bins of 0.3 counts~s$^{-1}$ width.  See text for details of
corresponding fit parameters.}
\end{center}
\label{fig:lagprob}
\end{figure*}
Previous observations of NGC~4051 with {\it EXOSAT} \cite{pl} and recently
{\it ASCA} \cite{gu}, indicate that the photon index of the
X-ray power-law is positively correlated
with the source luminosity, in the sense that the continuum becomes softer
as the source flux increases.  An inspection of the scaled
{\it EUVE} and {\it RXTE} lightcurves
indicates that this may indeed be the case; there is a tendency
for the linearly scaled {\it EUVE} flux to exceed the 2--10~keV count
rate when the source is bright, and vice versa when the source is dim. 
If the {\it EUVE} count rate can be described as a simple function of
the {\it RXTE} count rate, we can fit such a function to the
data.
The scaled lightcurves indicate that the {\it
EUVE} count rate scales with the {\it RXTE} count rate in a non-linear
way.  We suggest a simple function of the form:
\[
R_{\rm EUVE}=A\:R_{\rm RXTE}^{\:n}+C
\]
where $R_{\rm EUVE}$ and $R_{\rm RXTE}$ are the predicted {\it EUVE}
count rate and the {\it
RXTE} (4--10~keV, excluding 5--7~keV) count rate respectively. 
$A$, $n$ and $C$ are constants. 
The error on the predicted values
is given by:
\[
\Delta R_{\rm EUVE}=(R_{\rm EUVE}-C)\: \frac{n\:
\Delta R_{\rm RXTE}}{R_{\rm RXTE}}
\]
where $\Delta R_{\rm EUVE}$ and $\Delta R_{\rm RXTE}$ are the errors on
the {\it EUVE} and 4--10~keV count rates respectively.

For given parameters $A$, $n$ and $C$  we can calculate the predicted
value (and error) of $R_{\rm EUVE}$ for
each point in the {\it RXTE} lightcurve and compare with the actual
values by calculating a $\chi ^{2}$ value, defined as:
\[
\chi ^{2}=\sum^{n}\frac{(R_{\rm EUVE,measured}-R_{\rm EUVE})^{2}}
{\Delta R_{\rm EUVE,measured}^{2}+\Delta R_{\rm EUVE}^{2}}
\]
Where $R_{\rm EUVE,measured}$ and $\Delta R_{\rm EUVE,measured}$ are the
measured values of the {\it EUVE} count rate and associated error
respectively, and the sum is over the $n$ pairs of data points that are
measured simultaneously in both bands.
By stepping through a range
of values of the parameters we can attempt to find the set of best-fitting
model parameters. 
   
We use the 4--10~keV band of the {\it RXTE} data (excluding 5--7~keV), so that
we sample only the continuum component of the X-ray spectrum.  Due to
orbital constraints there are gaps in the {\it EUVE} lightcurve so that
not all {\it RXTE} data points have corresponding simultaneous {\it
EUVE} data points.  We therefore bin the lightcurves into identical 1~ks
bins (so the corresponding time for each bin is the same for both the
{\it EUVE} and 4--10~keV lightcurves).  In this way we only sample times
where the {\it EUVE} and {\it RXTE} data are simultaneous to within
1~ks.  This restriction is necessary because large flux changes can occur on
timescales of a few ks, so the flux in both bands must be measured as
simultaneously as possible to allow a good comparison between bands. 
We next fit the simultaneous data with our model.

We first fit a simple linear model to the data ($n=1$).  We
obtain best-fit values of $A=0.029\pm^{0.005}_{0.004}$ and 
$C=-0.006\pm^{0.005}_{0.007}$, with $\chi ^{2}=35.14$ for 31
degrees of freedom.  This model is acceptable, but a non-linear model
(allowing $n$ to be a free
parameter) improves the fit ($\chi ^{2}/d.o.f.=30.00/30$), with best-fit
values $A=0.012\pm^{0.013}_{0.006}$, $n=1.68\pm0.55$ and
$C=0.006\pm^{0.007}_{0.011}$.  Negative and positive values
of $C$ (as given by the linear and non-linear model fits) correspond to
constant flux components in the {\it RXTE} and {\it EUVE} bands
respectively.  Recently NGC~4051 was observed in an ultra-low state by
{\it BeppoSAX}, {\it RXTE} and {\it EUVE}
(Guainazzi et al. 1998, Uttley et al. 1999). 
The {\it BeppoSAX} and {\it EUVE}
observations revealed a constant soft component which may be associated
with the extended emission in the host galaxy imaged by {\it ROSAT} (Singh
1999).  This
component contributes an EUV flux consistent with the constant value
estimated by our non-linear model fit.  Additionally, {\it RXTE} and {\it
BeppoSAX} detected a constant hard component which may be due to
reflection from a molecular torus.  On its own, such a component would
lead to a negative value of $C$.  The exact
value of $C$, which is at present not well constrained, is
determined by a trade-off between these two
constant components.  We note also that fluctuations in the cosmic X-ray
background, which are unaccounted for by the PCA background model, could
lead to some additional constant negative or positive offset which will
contribute to $C$.  However,
the good agreement between the simultaneous medium-energy X-ray spectra
obtained by {\it BeppoSAX} and {\it RXTE} during the ultra-low state
(Uttley et al. 1999)
shows that any offset due to inaccurate background estimation is much 
smaller than any offset due to the torus component.  Using our data, we
cannot formally rule out either the
non-linear or linear model, however the $F$-test indicates that
the non-linear model is better at describing the data at $63\%$
confidence.  In figure 4 we show a comparison of the linear and non-linear
models with the data.

The fact that we can successfully model (reduced $\chi ^{2}=1.0$)
the relationship between both energy bands to a simultaneous time
resolution of 1~ks implies that there is no significant lag
between the bands (i.e. any lag is less than 1~ks).  This result is
consistent with the less-than-unity value of the zero-lag peak
in the lower resolution CCF.  We can rule out
the possibility that our model is so general that it will fit any time lag by
shifting the lightcurves with respect to each other and fitting the
model.  If we cause the {\it EUVE} lightcurve to lag the {\it RXTE}
lightcurve by 1~ks, the best-fitting non-linear model yields
$\chi ^{2}/d.o.f.=65.78/23$.  If we cause the {\it RXTE} lightcurve to
lag the {\it EUVE} lightcurve by 1~ks we obtain
$\chi ^{2}/d.o.f.=39.38/35$.  If the {\it RXTE} lightcurve lags the {\it
EUVE} lightcurve by 2~ks we find $\chi ^{2}/d.o.f.=60.78/38$.  Clearly
the model does not easily fit other lags, although a lag of 1~ks
between the X-ray and EUV bands is allowed by the model.  We might
expect the model to fit short lags even if the true lag is zero,
since both lightcurves are autocorrelated on short timescales
($\sim$ks).  In any case, it remains true that effectively zero lag
between the bands is adequate to describe the data.  

Recent simultaneous
observations of NGC~5548 with {\it RXTE} and {\it EUVE} showed evidence
that the 2--20~keV X-ray variations lagged changes in the EUV by 35~ks
(Chiang et al. 1999). 
Scaling this lag by the black hole mass, using the masses of NGC~4051
($1.4\times10^{6}$~M$_{\odot}$) and NGC~5548 ($8\times10^{7}$~M$_{\odot}$)
measured by reverberation mapping (Wandel, Peterson \& Malkan 1999), we
predict an X-ray--EUV lag of $\sim600$~s in NGC~4051, consistent with
our upper limit. 
  
\section{Discussion}
We have shown that the variable X-ray and EUV fluxes almost certainly come from
the same spectral component, and that the best-fitting model to
describe the relationship between the fluxes in both bands is probably
non-linear, implying that increases in continuum flux are accompanied by
spectral slope
changes.  Specifically, the power-law steepens as the continuum flux
increases.  This behaviour was observed previously by {\it ASCA}
\cite{gu}.  Using {\sc pimms} we can estimate the change in power-law
slope associated with a doubling of the {\it RXTE} count rate assuming
$n=1.68$.  The corresponding slope increase is $\sim0.1$, a value consistent
with the results obtained by Guainazzi et al. 

By constraining the lag between the EUV and 4--10~keV bands to be less
than 20~ks or less than 1~ks (depending on whether
we take the conservative
result of the ZDCF, or the more speculative result of the scaling
model), we can constrain the physical size of the Comptonising region in
the context of simple upscattering models.  In this context, the lag between
the EUV and 4--10 keV bands corresponds to the
time taken to upscatter EUV photons
to medium X-ray energies (regardless of the energy of the initial
seed photons).  We now consider a simple Comptonisation model, where the
EUV photons originate at the centre of a homogeneous spherical cloud of thermal
electrons, and are upscattered on their way through the cloud to produce the
medium-energy X-ray photons.  We use this simple model to estimate an upper
limit to the size of the X-ray emitting region. 

The time spent to upscatter a photon
from an energy $E_{\rm 1}$ to a higher energy $E_{\rm 2}$ is:
\begin{equation}
t_{\rm up}=\frac{N\: \lambda }{c}     
\end{equation}      
where $N$ is the number of scatterings required to raise the energy from
$E_1$ to
$E_2$ and $\lambda$ is the mean free path of the photon between
scatterings.  The mean free path is given by:
\begin{equation}
\lambda = \frac{1}{n_{\rm e}\: \sigma _{\rm T}}
\end{equation}
where $n_{\rm e}$ is the electron density and $\sigma _{\rm T}$ is the Thomson
cross-section (assuming $E_1 \ll m_{\rm e}c^2$). 
We can express $n_{\rm e}$ in terms of the optical depth of
the Comptonising cloud, $\tau$ and the radius of the cloud, $R$:
\begin{equation}
n_{\rm e}=\frac{\tau}{\sigma _{\rm T}\:R}.
\end{equation}
Incorporating equations 2 and 3 into equation 1, and rearranging
yields:
\begin{equation}
R=\frac{c\:\tau \:t_{\rm up}}{N}
\end{equation}
so that substituting our maximum lag for $t_{\rm up}$ sets an upper
limit on $R$.  The number of collisions required to scatter the photons
from energy $E_{\rm 1}$ to $E_{\rm 2}$ depends on the electron temperature,
$T_{\rm e}$ (e.g. see Longair, 1992).  In the non-relativistic regime,
the number of scatterings is given by:
\begin{equation}
N=\frac{{\rm log}\:(E_{\rm 2}/E_{\rm 1})}{{\rm log}\:(1+(4\:k\:T_{\rm
e}/m_{\rm e}\:c^2))}.
\end{equation}
The electron temperature is not well constrained, but
because the power-law continuum extends right across the useful PCA band
(up to 15 keV) with no cut-off at the highest energies, we are justified
in assuming a lower limit to the electron temperature of $\sim25$~keV. 
OSSE observations of Seyfert galaxies show high energy cutoffs in
their continua which correspond to typical electron temperatures of
$\sim100$~keV ($\sim60$~keV in the case of NGC~4151) \cite{zd},
consistent with our simple non-relativistic assumption.

The ratio of photon energies between the two bands is $> 20$, so for an
example where
$k\:T_{\rm e}=25$~keV, $N>16$.  For $k\:T_{\rm e}=100$~keV, $N>5$. 
For photon energies $E \ll k\:T_{\rm e}$, the optical depth for Compton
scattering in a spherical cloud can be determined from the spectral
slope of the resulting Comptonised continuum \cite{po}:
\[
\tau = \left(\frac{\pi^2}{4\:\left[(\alpha+3/2)^2-9/4 \right]}\:\frac{m_{\rm
e}\:c^2}{k\:T_{\rm e}} \right)^\frac{1}{2} -0.5
\]
where $\alpha$ is the energy spectral index of the Comptonised continuum.  Using
the spectral index of 1.3 measured by {\it RXTE}, we find $\tau =
1$ for $k\:T_{\rm e}=100$~keV, and $\tau = 2.5$ for $k\:T_{\rm e}$ of
25~keV.  Therefore, in the range of temperatures 25--100~keV, we expect
the ratio $\tau /N < 0.2$.  Using this limit in
equation 4, we arrive at an upper limit for $R$ assuming $t_{\rm up}<1000$~s:
\[
R<6\times10^{12}\;\;{\rm cm}.
\]
A size of $6\times10^{12}$~cm corresponds to $\sim 20$
Schwarzschild radii for a $10^6$~M$_\odot$ black hole.  Thus, if the
power-law continuum is produced by a single central source, the emitting
region is very
close to the central black hole.  Alternatively, in disk corona models
(e.g. see Haardt, Maraschi \& Ghisellini, 1997)
the power-law continuum is produced by the scattering of thermal
photons from an accretion disk in a hot corona above the disk.  In this
case, the continuum may originate in many scattering regions spread over
the disk, so the size constraint we
impose may represent the size of a typical scattering region. 
Note that this size limit increases by a factor $\sim20$ if we instead
choose the more conservative 20~ks upper limit on the lag,
inferred from the ZDCF.

We note here that Monte Carlo
simulations of thermal Comptonisation in the relativistic regime
\cite{sk} show that if $\alpha >1$, $\tau<0.1$ for $k\:T_{\rm
e}>250$~keV.  In the case of very low optical depth to scattering, EUV and
X-ray photons may undergo only one upscattering from the original
seed photons before leaving the Comptonising cloud.  In this
case, upper limits to lags between the two bands yield no
information regarding the size of the Comptonising region.  This
situation is extreme however, but in the case of NGC~4051 we cannot be
completely confident in our assertion of a small Comptonising region
until finite lags are measured between different energy
bands, or a spectral cut-off is determined. 

Finally, we note that although our constraint on the size of the
emitting region in NGC~4051 is model dependent, simple causal arguments
place a model-independent upper limit of $3\times10^{13}$~cm on the
separation between the EUV and X-ray emitting regions, if the lag
between both bands is less than 1~ks.  Combining this
constraint with the rapid variability seen in both bands, we can infer that the
continuum in both bands is emitted co-spatially within a region of that
size (i.e., 1000 light-seconds).

\section{Conclusions}
In the introduction to this paper, we raised two key questions which
would shed light on the nature of the X-ray continuum production
mechanism in Seyfert Galaxies, namely {\it how low in energy does the
power-law extend?} And {\it are there any lags between the X-ray band
and lower energy bands?} 
We have answered these questions to some extent in the case of
NGC~4051.  We have shown that the EUV variability of NGC~4051 can be 
adequately explained as being due to the extension of the X-ray
power-law into the EUV band.  This implies that the source of `seed'
photons required must be looked for at lower energies.  Done et al.
(1990) have shown that NGC~4051 does not vary significantly in the
optical band, so the X-ray power-law must end somewhere in the UV, Far UV
or in the unobservable, heavily absorbed EUV band above the Lyman edge. 

We have shown that there are no significant lags down to timescales of
20~ks, and that if the EUV emission is the extension of the X-ray
power-law (of varying slope or otherwise), the lag is reduced much further
to less than 1~ks.  This places strong constraints on the size of the
Comptonising region, implying that it lies within a few gravitational
radii of the central black hole for a typical black hole mass of
between $10^6$ and $10^7$~M$_\odot$, if the X-ray source is central.

Since NGC~4051 is a low luminosity Seyfert galaxy, we might expect that
more luminous AGN will contain larger continuum emission regions. 
The lack of strong variability in higher luminosity AGN
makes the type of study outlined in this paper difficult, but the high
count rate, high $S/N$
observations that will be available with {\it XMM} should prove
excellent for constraining the size of the continuum emission region
across a range of source luminosities.

\subsection*{Acknowledgments}
We wish to thank the {\it RXTE} and {\it EUVE} schedulers and Guest
Observer support teams for efficiently co-ordinating and supporting
these observations.  We also thank the anonymous referee for helpful 
suggestions. 
PU acknowledges financial support from the Particle Physics and
Astronomy Research Council, who also provided grant support to IM$^{\rm c}$H. 
AF was supported by AXAF Science Center NASA contract NAS 8-39073.  IC
was supported by NASA grants NAG5-3191 and NAG5-3174.

\bsp

\begin{thebibliography}{99}
\bibitem[\protect\citename{Abbot et al. }%
1996]{ab} Abbott M., Boyd W., Jelinsky P., Christian C., Miller-Bagwell
A., Lampton M., Malina R. F., Vallerga J. V., 1996, ApJS, 107, 451
\bibitem[\protect\citename{Alexander }%
1997]{al} Alexander T., 1997, in Maoz D. et al., ed., Astronomical Time
Series, Kluwer Academic Publishers, Netherlands, p. 163
\bibitem[\protect\citename{Chiang et al. }%
1999]{chiang} Chiang J., Reynolds C. S., Blaes O. M., Nowak M. A.,
Murray N., Madejski G., Marshall H. L., Magdziarz P., 1999, ApJ, submitted
\bibitem[\protect\citename{Clavel et al. }%
1992]{cl} Clavel J. et al., 1992, ApJ, 393, 113
\bibitem[\protect\citename{Done et al. }%
1990]{do} Done C., Ward M. J., Fabian A. C., Kunieda H., Tsuruta S.,
Lawrence A., Smith M. G., Wamsteker W., 1990, MNRAS, 243, 713
\bibitem[\protect\citename{Edelson \& Krolik }%
1988]{ek} Edelson R. A., Krolik J. H., 1988, ApJ, 333, 646
\bibitem[\protect\citename{Edelson \& Nandra }%
1999]{en} Edelson R. A., Nandra K., 1999, ApJ, 514, 682
\bibitem[\protect\citename{Edelson et al. }%
1996]{edet96} Edelson R. A. et al., 1996, ApJ, 470, 364
\bibitem[\protect\citename{Elvis, Lockman \& Wilkes }%
1989]{elv} Elvis M., Lockman F. J., Wilkes B. J., 1989, AJ, 97, 777
\bibitem[\protect\citename{Fiore et al. }%
1992]{fi} Fiore F., Perola G. C., Matsuoka M., Yamauchi M., Piro L.,
1992, A\&A, 262, 37
\bibitem[\protect\citename{Fruscione et al. }%
1999]{fr} Fruscione A., Cagnoni I., M${\rm ^{c}}$Hardy I. M., Papadakis I. E.,
1999, in prep.
\bibitem[\protect\citename{Guainazzi et al. }%
1996]{gu} Guainazzi M., Mihara T., Otani C., Matsuoka M., 1996, PASJ,
48, 781
\bibitem[\protect\citename{Guainazzi et al. }%
1998]{gu98} Guainazzi M. et al., 1998, MNRAS, in press 
\bibitem[\protect\citename{Haardt, Maraschi \& Ghisellini }%
1997]{hmg} Haardt F., Maraschi L., Ghisellini G., 1997, ApJ, 476, 620
\bibitem[\protect\citename{Lawrence et al. }%
1987]{law} Lawrence A., Watson M. G., Pounds K. A., Elvis M., 1987, Nat,
325, 696
\bibitem[\protect\citename{Leighly et al. }%
1996]{lei} Leighly K. M., Mushotzky R. F., Yaqoob T., Kunieda H.,
Edelson R., 1996, ApJ, 469, 147
\bibitem[\protect\citename{Longair }%
1994]{lon} Longair M. S., 1992, High Energy Astrophysics Vol. 1, CUP,
Cambridge
\bibitem[\protect\citename{M${\rm ^{c}}$Hardy }%
1988]{mch88} M${\rm ^{c}}$Hardy I. M., 1988, Mem. Soc. Astron. Ital.,
59, 239
\bibitem[\protect\citename{M${\rm ^{c}}$Hardy et al. }%
1995]{mch} M${\rm ^{c}}$Hardy I. M., Green A. R., Done C., Puchnarewicz E. M.,
Mason K. O., 

Branduardi-Raymont G., Jones M. H., 1995, MNRAS, 273, 549
\bibitem[\protect\citename{Marshall et al. }%
1997]{mar} Marshall et al., 1997, ApJ, 479, 222
\bibitem[\protect\citename{Nandra et al. }%
1998]{nanetal} Nandra K., Clavel J., Edelson R. A., George I. M., Malkan
M. A., Mushotzky R. F., Peterson B. M., Turner T. J., 1998, ApJ, 505, 594
\bibitem[\protect\citename{Nowak et al. }%
1999]{now} Nowak M. A., Vaughan B. A., Wilms J., Dove J. B., Begelman M.
C., 1999, ApJ, 510, 874
\bibitem[\protect\citename{Page }%
1985]{page} Page C. G., 1985, Sp. Sci. Rev., 40, 387
\bibitem[\protect\citename{Papadakis \& Lawrence }%
1995]{pl} Papadakis I. E., Lawrence A., 1995, MNRAS, 272, 161
\bibitem[\protect\citename{Posdnyakov, Sobol \& Sunyaev }%
1983]{po} Posdnyakov L. A., Sobol I. M., Sunyaev R. A., 1983,
Astrophysics \& Space Physics Reviews, 2, 263
\bibitem[\protect\citename{Ramos et al. }%
1997]{ra} Ramos E., Kafatos M., Fruscione A, Bruhweiler F. C.,
M${\rm ^{c}}$Hardy I. M., Hartman R. C., Titarchuk L. G., von Montigny C., 1997,
ApJ, 482, 167
\bibitem[\protect\citename{Singh }%
1999]{sin} Singh K. P., 1999, MNRAS, in press
\bibitem[\protect\citename{Sirk et al. }%
1997]{si} Sirk M. M., Vallerga J. V., Finley D. S., Jelinsky P., Malina
R.F., 1997, ApJS, 110, 347
\bibitem[\protect\citename{Skibo et al. }%
1995]{sk} Skibo J. G., Dermer C. D., Ramaty R., McKinley J. M., 1995,
ApJ, 446, 86
\bibitem[\protect\citename{Wandel, Peterson \& Malkan }%
1999]{wa} Wandel A., Peterson B. M., Malkan M. A., 1999, ApJ, in press
\bibitem[\protect\citename{Welsh et al. }%
1990]{we} Welsh B. Y., Jelinsky P., Vallerga J. V., Vedder P. W., Finley
D. S., Malina R. F., 1990, Proc. SPIE, 1343, 166
\bibitem[\protect\citename{Zdziarski et al. }%
1997]{zd} Zdziarski A. A., Johnson W. N., Poutanen J., Magdziarz P.,
Gierlinski M., 1997, Proc. of the 2nd INTEGRAL Workshop, ESA SP-382, 373
\end{thebibliography}
\end{document}